%% file: template.tex
\PassOptionsToPackage{table,dvipsnames,svgnames}{xcolor}

\documentclass[preprint]{vgtc}               % preprint
\usepackage{xcolor}

\makeatletter
\@ifpackageloaded{xcolor}{}{\usepackage{xcolor}}
\@ifundefined{color@NavyBlue}{\definecolor{NavyBlue}{rgb}{0,0,0.5}}{}
\makeatother

\onlineid{1431}

\preprinttext{Authors' version. To appear in IEEE Transactions on Visualization and Computer Graphics.}

\vgtccategory{Research}

\vgtcpapertype{please specify}

\title{Behavioral and Symbolic Fillers as Delay Mitigation for Embodied Conversational Agents in Virtual Reality}

\author{%
  \authororcid{Denmar Mojan Gonzales}{0009-0003-2336-232X},
  \authororcid{Snehanjali Kalamkar}{0000-0003-4818-0197},
  \authororcid{Sophie Jörg}{0000-0002-7910-8553}, and 
  \authororcid{Jens Grubert}{0000-0002-3858-2961}
}

\authorfooter{
  \item
  	Denmar Mojan Gonzales, Snehanjali Kalamkar, Jens Grubert are with Coburg University of Applied Sciences, Germany.
  	E-mail: denmar-mojan.gonzales@stud.hs-coburg.de, {snehanjali.kalamkar, jens.grubert}@hs-coburg.de.
  \item
  	Sophie Jörg is with the University of Bamberg.
  	E-mail: sophie.joerg@uni-bamberg.de.
}

\abstract{%
When communicating with embodied conversational agents (ECAs) in virtual reality, there might be delays in the responses of the agents lasting several seconds, for example, due to more extensive computations of the answers when large language models are used. Such delays might lead to unnatural or frustrating interactions. 
In this paper, we investigate filler types to mitigate these effects and lead to a more positive experience and perception of the agent. In a within-subject study, we asked 24 participants to communicate with ECAs in virtual reality, comparing four strategies displayed during the delays: a multimodal behavioral filler consisting of conversational and gestural fillers, a base condition with only idle motions, and two symbolic indicators with progress bars, one embedded as a badge on the agent, the other one external and visualized as a thinking bubble.  
Our results indicate that the behavioral filler improved perceived response time, three subscales of presence, humanlikeness, and naturalness. Participants looked away from the face more often when symbolic indicators were displayed, but the visualizations did not lead to a more positive impression of the agent or to increased presence. The majority of participants preferred the behavioral fillers, only 12.5\% and 4.2\% favored the symbolic embedded and external conditions, respectively.
}


\usepackage{subcaption}
\input{teaser}
\graphicspath{{figs/}{figures/}{pictures/}{images/}{./}} % where to search for the images

\usepackage{tabu}                      % only used for the table example
\usepackage{booktabs}                  % only used for the table example
\usepackage{lipsum}                    % used to generate placeholder text
\usepackage{mwe}                       % used to generate placeholder figures

\usepackage{mathptmx}                  % use matching math font
\usepackage{soul}
\usepackage{todonotes}
\usepackage{color}
\usepackage{colortbl}
\usepackage{tabularx}
\usepackage{graphicx}
\usepackage{multirow}
\usepackage{balance}
\usepackage{makecell}
\usepackage[final]{changes} %  [draft] to show changes or [final] to hide changes

\begin{document}

%%%%%%%%%%%%%%%%%%%%%%%%%%%%%%%%%%%%%%%%%%%%%%%%%%%%%%%%%%%%%%%%
%%%%%%%%%%%%%%%%%%%%%% START OF THE PAPER %%%%%%%%%%%%%%%%%%%%%%
%%%%%%%%%%%%%%%%%%%%%%%%%%%%%%%%%%%%%%%%%%%%%%%%%%%%%%%%%%%%%%%%

\firstsection{Introduction}

\maketitle

\input{sections/01_introduction}

\input{sections/02_relatedWork}

\input{sections/03_userStudy}
\input{sections/04_results}

\input{sections/05_discussion}

\input{sections/06_conclusion}

% The authors wish to thank A, B, and C. This work was supported in part by
% a grant from XYZ.}

\balance

\bibliographystyle{abbrv-doi}

\bibliography{template}

% \begin{comment}

% \appendix % You can use the `hideappendix` class option to skip everything after \appendix

% \section{About Appendices}
% Refer to \cref{sec:appendices_inst} for instructions regarding appendices.

% \section{Troubleshooting}
% \label{appendix:troubleshooting}

% \subsection{ifpdf error}

% If you receive compilation errors along the lines of \texttt{Package ifpdf Error: Name clash, \textbackslash ifpdf is already defined} then please add a new line \verb|\let\ifpdf\relax| right after the \verb|\documentclass[journal]{vgtc}| call.
% Note that your error is due to packages you use that define \verb|\ifpdf| which is obsolete (the result is that \verb|\ifpdf| is defined twice); these packages should be changed to use \verb|ifpdf| package instead.

% \subsection{\texttt{pdfendlink} error}

% Occasionally (for some \LaTeX\ distributions) this hyper-linked bib\TeX\ style may lead to \textbf{compilation errors} (\texttt{pdfendlink ended up in different nesting level ...}) if a reference entry is broken across two pages (due to a bug in \verb|hyperref|).
% In this case, make sure you have the latest version of the \verb|hyperref| package (i.e.\ update your \LaTeX\ installation/packages) or, alternatively, revert back to \verb|\bibliographystyle{abbrv-doi}| (at the expense of removing hyperlinks from the bibliography) and try \verb|\bibliographystyle{abbrv-doi-hyperref}| again after some more editing.
% \end{comment}

\end{document}

%% file: teaser.tex
\teaser{
  \centering
  \begin{subfigure}[t]{0.24\linewidth}
    \includegraphics[width=\linewidth]{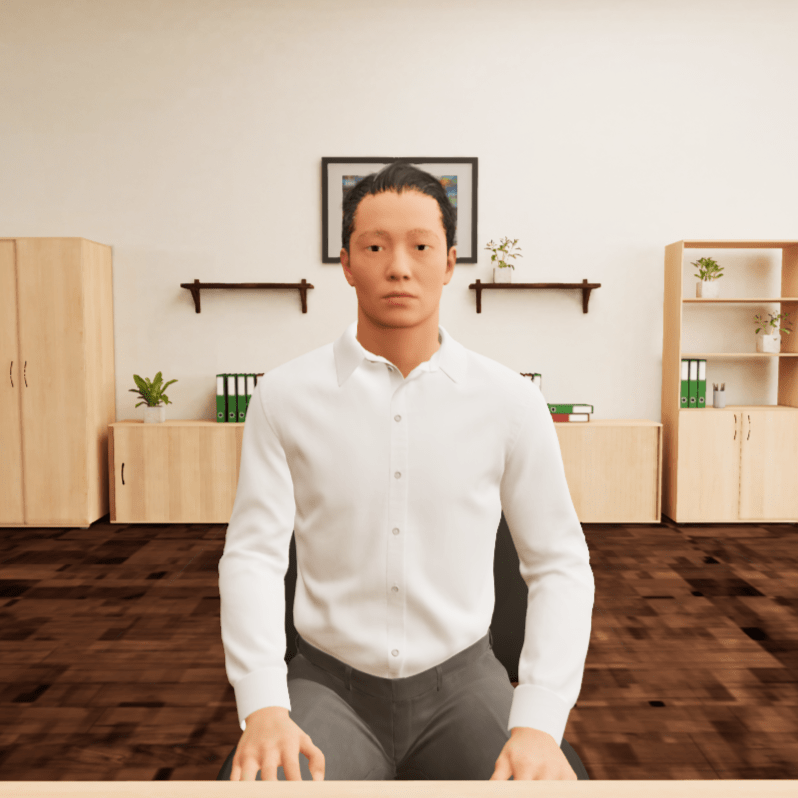}
    \caption*{(a)}
  \end{subfigure}
  \hfill
  \begin{subfigure}[t]{0.24\linewidth}
    \includegraphics[width=\linewidth]{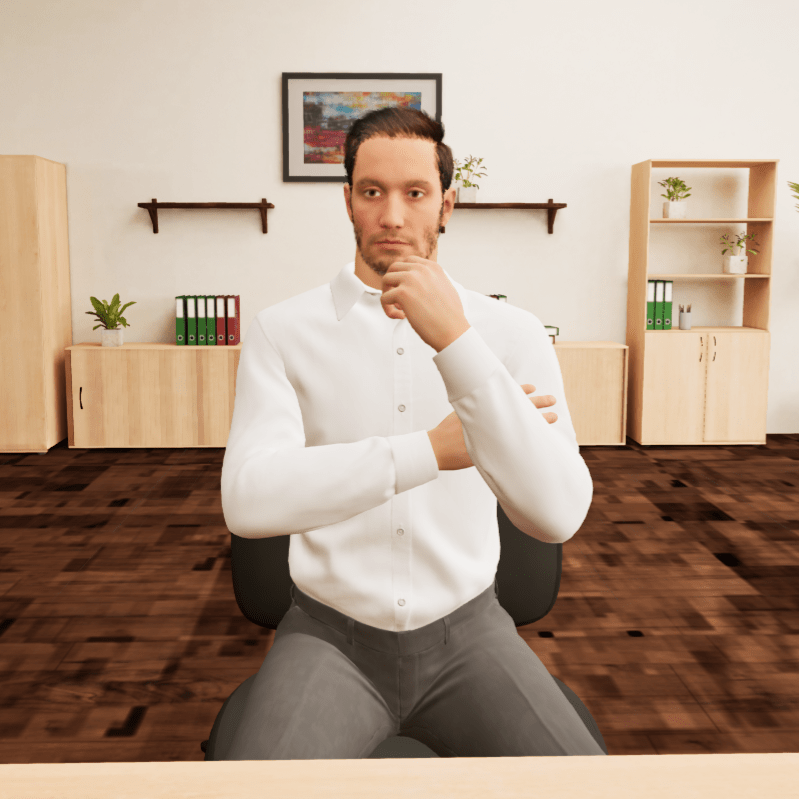}
    \caption*{(b)}
  \end{subfigure}
  \hfill
  \begin{subfigure}[t]{0.24\linewidth}
    \includegraphics[width=\linewidth]{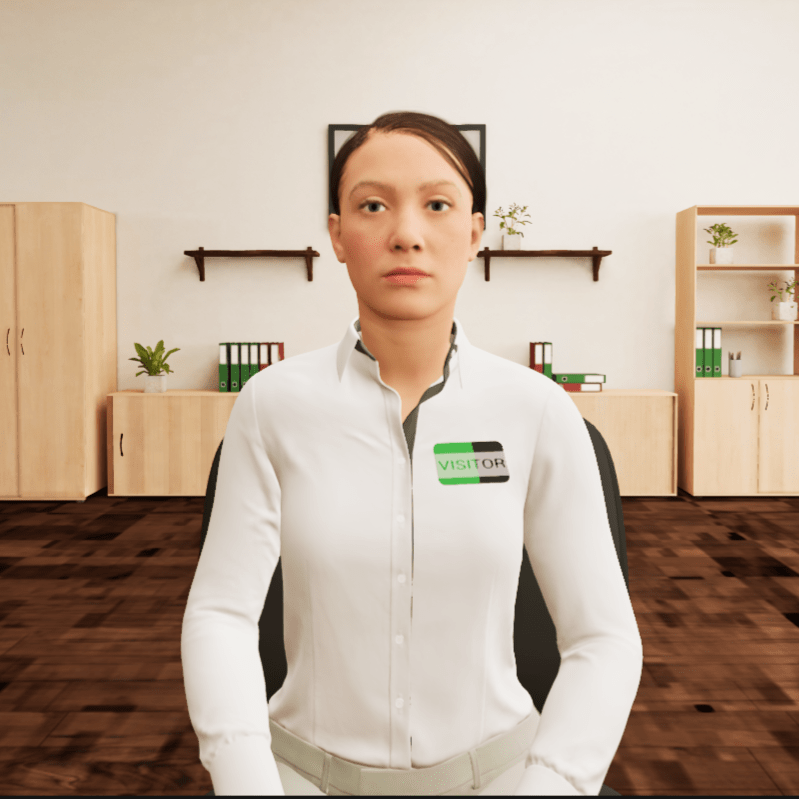}
    \caption*{(c)}
  \end{subfigure}
  \hfill
  \begin{subfigure}[t]{0.24\linewidth}
    \includegraphics[width=\linewidth]{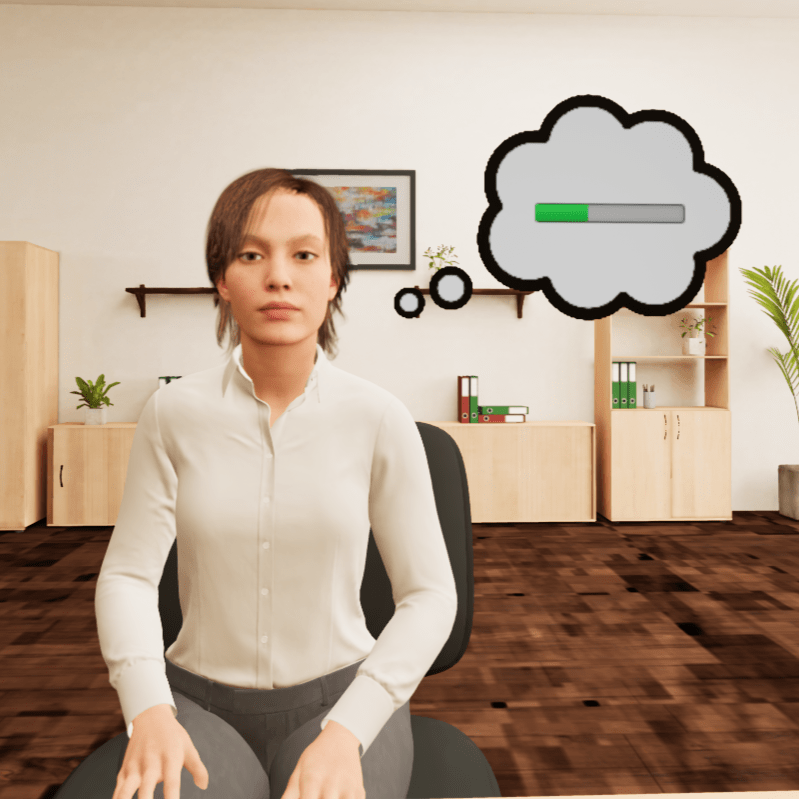}
    \caption*{(d)}
  \end{subfigure}
  \caption{Visualization of response delay mitigation strategies for embodied conversational agents (ECA): a) \textsc{base} condition where the ECA continues playing idle animations during a "thinking" phase; b) \textsc{behavioral} condition in which the ECA performs thinking animations (e.g., gaze aversion, filler gestures) to signal cognitive processing; c) \textsc{embedded} condition where a visual loading indicator in the form of a green progress bar lights up across the ECA's visitor badge during the thinking phase, culminating in the ECA’s response once fully filled; d) \textsc{external} condition featuring a floating thinking bubble next to the ECA, containing a loading bar that gradually fills up until the \replaced{ECA}{VA} delivers a response.}
  \label{fig:teaser}
}

%% file: sections/01_introduction.tex
Embodied conversational agents (ECAs)  are computer-generated characters "that demonstrate many of the same properties as humans in face-to-face conversation, including the ability to produce and respond to verbal and nonverbal communication" \cite{cassell2001embodied} and are frequently used as part of many virtual reality (VR) experiences. 

Advancements in Generative Artificial Intelligence (GenAI), \replaced{large language models }{large-language-models} (LLMs) and \replaced{vision language models }{vision-language-models} (VLMs) in particular have led to a growing interest in combining ECAs with GenAI to create more intelligent behaviors than early virtual agents (e.g., \cite{maslych2024takeaways, schmidt2020intelligent, schmidt2024natural,  sonlu2021conversational, yang2024effects, zhu2023free}).

However, communicating with an ECA can induce response delays of the agents lasting several seconds, for example, due to more extensive computations of the answers using LLMs, impacting fluidity in the conversation. Such extensive delays typically do not happen when communicating with humans and might lead to unnatural or frustrating interactions. Several strategies have been developed to mask these response delays, the two main ones being the addition of conversational fillers (e.g., "um", "uh") and behavioral fillers (e.g., changing posture, looking away, scratching one's head).

While these delay mitigation strategies aim to imitate the behavior of real people, at the current state of the technology, they do not fool people into believing that their counterpart is a real person. Repetitive behavior or conversational fillers might, on the contrary, be irritating over time and, depending on the context, can negatively impact the interaction between a human and an agent \cite{Jeong19conv, liu2024hmm}. Furthermore, such fillers do not give users an idea of the time a response might take. 
% not super happy with sentences, but trying to write thoughts down
One advantage of VR applications is that we can give the user additional information in ways that are not supported by reality. 
%\textcolor{blue}{In this regard, multimodal feedback, including visual feedback, along with the conversational and behavioral fillers, has been shown to improve the sense of presence and immersion \cite{elfleet2024investigating}. However, this visual indicator was a spinner icon that indicated the processing stage of the virtual agent.} %SJ: I don't think that reasoning works as that was in combination with conversational and behavioral
Progress indicators have been used in a variety of applications and have been found to be beneficial for user experiences \cite{myers1985importance}.
So, in addition to simulating the behavior of real humans thinking about an answer, we explore the introduction of visual information that indicates that the application is making progress and gives the user an idea of when an answer can be expected. Expected waiting times might be estimated, for example, based on internet speed, context, or question type. 
%\sj{Can we actually broadly estimate the time an LLM might need to answer? Are there maybe references about how to do that?} 

%In this paper, we investigate if visualizations close to the virtual character can mitigate these effects and lead to a more positive experience and perception of the virtual character. 
Our contribution lies in evaluating the effects of a multimodal behavioral filler as well as of symbolic filler visualizations (with progress indicators) on the perceived response time, presence, and the users' impression of the agent. %Furthermore, we explore the use of fillers with progress indicators. 
To this aim, we asked participants to communicate with ECAs, comparing four visual delay mitigation strategies. Specifically, we compared thinking-like animations (\textsc{behavioral} condition) with external symbolic (\textsc{external}) and embedded symbolic (\textsc{embedded}) visualization and no visualization (\textsc{base}) in a user study with 24 participants in VR. Our results indicate, amongst others, that the \textsc{behavioral} condition was the most preferred (67\%), and that the response time was perceived as significantly more appropriate than in the other conditions.

%% file: sections/02_relatedWork.tex
\section{Related Work}

% Structure:
% 1. Virtual humans, agents, avatars, conversational agents: definitions usage - maybe some of that can go in the intro
% Perception of such VH, agents, conversations, turn-taking cues
% 2. Conversational agents with LLMs, AI... Latency occurs
% 3. Previous work on latency mitigation strategies
% 4. Visualizations in VR/AR

%\subsection{Perception of Embodied Conversational agents}

While intelligent virtual agents can communicate via text or voice, embodied conversational agents have a visual representation such as a fully animated body and face \cite{Kim18}. Such embodied agents can be used in many types of applications and fields including healthcare, education, or social VR \cite{Wan24LLMagents,Lupetti23health}.
Creating embodied conversational agents has been a topic of research for many years with a focus on synthesizing humanlike verbal and nonverbal behavior \cite{Norouzi18survey}. Conversations between people are a complex interaction of signals including vocal cues, body language, facial expression, gaze, or proxemics. Small changes or unnatural inaccuracies in conversation with virtual characters or embodied agents can affect how a conversation or agent is perceived \cite{Adkins23hands, Ehret21prosody}. 

When humans communicate with each other, pauses and silences are normal and expected behavior \cite{jaffe2001rhythms, mukawa2014verbal}. It has been shown that they can augment the perceived naturalness of interactions in conversational systems \cite{gnewuch2018faster}, but also that excessive pauses can negatively impact users' perception of agents \cite{yang2015effect} with preferred response times being reported to be around one second \cite{shiwa2008quickly}.

% conversational fillers

%\subsection{Delay Mitigation Strategies}

Prior work has investigated the use of non-lexical fillers (such as "um") and lexical fillers (such as "let me think") as delay mitigation strategies. For example, Gambino et al.~\cite{gambino2018testing} indicated that given an equal response time, users preferred conversational fillers over silently waiting for a response.
A study by Goble and Edwards showed that social presence increased when a robot spoke with vocal fillers \cite{Goble18vocalrobot}.
Boukaram et al.~\cite{boukaram2021mitigating} compared generic fillers with contextualized ones when seeing the face of a web-based agent. The agent's response time was judged to be more acceptable when using any type of filler in contrast to no filler, and perceived competence improved with contextualized fillers. However, the fillers had no significant effect on likeability. 
Conversational fillers can also have negative effects. Jeong et al.~\cite{Jeong19conv} showed that voice-based conversational agents using conversational fillers were perceived as less intelligent than when such fillers were not used.  Liu et al.~\cite{liu2024hmm} indicated that conversational fillers in a sales and promotion context can trigger consumer's suspicion of ulterior motives and lead to decreased purchase intentions.

Besides verbal feedback, non-verbal cues are also important to enhancing real-world interactions and interactions with virtual agents \cite{wang2021examining}. Latoschik et al.~\cite{latoschik2019not} highlighted the need for low latencies for
reliable communication of embodied social signals.
% conversational, gestural, multimodal fillers
Kum and Lee \cite{kum2022journal} contrasted the use of conversational fillers with gestural fillers in a screen-based interaction. They showed that gestural fillers could mitigate user-perceived latency more effectively than conversational ones while simultaneously increasing the perceived \added{behavioral naturalness, positive impression, and competence of the virtual agents, as well as the willingness to interact with the agents again}. 
Elfleet and Chollet \cite{elfleet2024investigating} investigated the effectiveness of multimodal feedback strategies (filled pauses, nonverbal turn-taking behaviors, and visual feedback) in mitigating perceived latency. In a user study (n=18), they compared a baseline condition (no feedback) with a feedback-enhanced condition. Their results indicated that the multimodal feedback condition significantly improved the sense of presence and immersion. At the same time, it is important that both verbal and non-verbal feedback match \cite{church1986mismatch}. In a pilot study (n=8), Maslych et al.~\cite{maslych2024takeaways} compared perceived avatar realism and responsiveness, user preference, and participants' gaze behavior for conversational agents with three delay types: lights indicating its status (idle, listening, thinking, speaking), a loading bar appearing during thinking status, and no feedback. Most participants preferred the status lights, but no significant effects between feedback conditions were found for realism and responsiveness.

In graphical user interfaces, symbolic indicators such as linear or circular progress bars are well established and are known to impact user satisfaction and perceived waiting times \cite{gronier2019does,myers1985importance}.  
Similar effects could be possible for embodied conversational agents that answer with delays due to computations with LLMs. Gnewuch et al.~showed that typing indicators increase the social presence of chatbots at least for novice users \cite{gnewuch2018chatbot}.

Complementary to prior work, we are focusing on investigating the effects of different symbolic feedback indicators (embedded in the agent and external to the agent) on the user experience.  We compare these feedback indicators to a base condition without any feedback and to a multimodal behavioral (gestural and conversational) filler.

%\sj{Is there general research about progress bars \cite{myers1985importance} or other indicators of waiting time, e.g, about user satisfaction when knowing how long a copying process takes?}
%\sj{Have there been studies using time indicators for other conversational agents, e.g, some using just voice or even text?}
%\cite{wintersberger2020tell} LESS RELEVANT: Using waiting time and queue indicators for filling wait time in . Our results, based on subjective waiting experience, service quality, as well as user experience and affect confirm that digital enterprises have to provide maximum transparency to customers, e.g., details about waiting time and queue status. In addition, they might use “time-fillers” in waiting situations to further improve customer satisfaction.

% maybe time indicators have been investigated for other IVAs such as Alexa or chatgpt

%% file: sections/03_userStudy.tex
\section{User Study}
%\subsection{Study Design}

We designed a user study to investigate the effects of different filler types on the user experience when interacting with an ECA. Specifically, we wanted to understand the differences between behavioral multimodal (gesture + speech) indicators of waiting time as studied in prior work and symbolic ones. 

As it is not straightforward to integrate symbolic indicators in or around a virtual agent without impacting the perceived naturalness of interaction with the agent, we designed two variations of linear progress indicators aiming at reducing this impact. 
\added{We specifically chose a progress bar for the symbolic indicator as it is well established and known to provide greater satisfaction to the users \cite{myers1985importance}.}
First, we integrated a linear progress bar in a thinking bubble visualization that would appear next to the agent when it is pausing to generate an answer as shown in \cref{fig:teaser}(d). 
\added{This idea was inspired by comics, where thinking action is indicated using a thinking bubble \cite{mccloud1993understanding}.}
While this visualization can be seen as a context-independent way to integrate a symbolic wait time indicator, it is also potentially distracting\added{, as it floats outside the agent}. 
Hence, we designed a progress bar visualization embedded in the virtual agent, in our case integrated in a badge on the chest of the character, see \cref{fig:teaser}(c),
\added{to have a minimalistic indicator relevant to our scenario, which kept the cue within the user’s primary focus.}

We conducted a within-subjects study with one independent variable \textsc{filler type} with four conditions visualized in \cref{fig:teaser}. In the \textsc{base} condition, there are no symbolic or behavioral fillers. The agent continues to play idle animations until it responds. In \textsc{behavioral} we combined animations with a non-lexical verbal filler ("Hhm") with six pitches. The verbal filler occurred every three questions. To avoid repeating the same motions, we created nine different animations. In the \textsc{external} condition, we displayed the progress bar integrated into the thinking bubble as described above. For the \textsc{embedded} condition, the progress bar was integrated into the badge as described above. 
% \sj{Is there are reason why 9 was highlighted? How many non-lexical fillers do we have? }
% \todo[inline]{D: The 9 was highlighted for me because I still had to fill it in. I just re-checked and we only have "Hhm" but in six different pitches picked randomly.}

As task, participants assumed the role of a Human Resources (HR) professional conducting four virtual job interviews in an office environment. 
\added{We chose this scenario as it provides a structured setting where the participants would be more sensitive to timing and motivated to maintain conversational flow. The formality of conducting interviews heightens the user’s awareness of interaction quality, which we deemed necessary for studying perceived responsiveness and naturalness.}
During the task, the participants were instructed to ask each agent ten predefined interview questions to assess the agents' qualifications and gather further insight.
\added{All ten questions were identical and presented in the same order across all conditions. Each question had four possible responses, which were randomized across conditions (and thus across agents) and not repeated between conditions.
As perception of the agents could be influenced by the subjective quality of their responses, all responses were designed to be neutral, professional, and as similar as possible.}
Between the user's questions and the agent's response, we introduced a pseudo-random waiting time consisting of three durations (following related work \cite{kum2022journal}): short (2s), medium (4s), and long (8s). The durations were distributed evenly, with the short duration occurring four times, and the medium and long durations occurring three times each, in a random order throughout the interview. 

\subsection{Measurements} 

We evaluated the impact of our filler types on perceived response time, presence, the impression of the agent, and user preferences. Specifically, our measures were as follows.

To assess the perceived response time, we asked participants to answer two questions on 7-point Likert scales: "The response time of the virtual human you just talked to was appropriate", which is similar to questions asked in previous related work \cite{boukaram2021mitigating,kum2022journal}, and "The virtual human you just talked to answered when expected", \added{defined by us to capture how well the participants could predict the agent’s response time based on the fillers shown.}

For presence, we followed The Temple Presence Inventory (TPI) \cite{Lombard09,Lombard11}, using the subscales parasocial interaction (we omitted the question on eye-contact), engagement, and social realism. Additionally, we evaluated the perceived social presence with four questions from Nowak and Biocca \cite{Nowak03}: "To what extent did you feel able to assess your partner’s reactions to what you said?", "To what extent was this like a face-to-face meeting?", "To what extent was this like you were in the same room with your partner?", and "To what extent did your partner seem ‘real’?".

Participants' impression of the agent was assessed on 7-point scales with questions from the Godspeed questionnaire and related work \cite{Bartneck09godspeed,elfleet2024investigating}.
Our constructs and scales in detail were humanlikeness (machinelike to humanlike, mechanical to organic, artificial to lifelike, and moving rigidly to moving elegantly), perceived intelligence (incompetent to competent, ignorant to knowledgeable, inert to interactive, and unintelligent to intelligent), likeability (awful to nice, dislike to like, unpleasant to pleasant, and unkind to kind), willingness to interact again ("Would you be willing to talk to the virtual human you just talked to next time?"), competence ("How likely is it that you would hire this person?"), naturalness (extremely unnatural to extremely natural), and assuredness (extremely eerie to extremely reassuring). % sj: better to make a table with the detailed questions?

For preference, we asked participants which person they would hire after they saw all four conditions. Finally, we inquired about the impact of each condition -- the symbolic time indicator "thinking bubble", the symbolic time indicator "visitor badge", the behavioral animations, or the absence of animations --  on the participants' experience.  
Participants were encouraged to write down any further comments about the different conditions and about their experience at the end of the experiment.

We complemented the subjective measures with an objective measure about the relative gaze time spent looking at the agent's face while it was generating a response. Gaze data was sampled at a frequency of 24 Hz. To obtain a smooth estimate of participants’ gaze aversion over time, we calculated a \replaced{cumulative average}{ rolling average} by summing the gaze deviation values and dividing by the number of samples. %This provided a continuous measure of average gaze deviation throughout the agent's response phase.
%\sj{We did also measure participants' gaze in different areas of interest, right? Do we want to mention that? We could, for example, check if participants looked at the badge in the embedded condition and if they looked at the bubble.}

%, as well as the average gaze deviation. To calculate gaze deviation, we defined the target point as the virtual agent’s nose bridge and computed the angular deviation between the participant’s gaze direction and the vector from the gaze origin to the target. This resulted in a scalar value representing the deviation angle in degrees for each sample.

%\cite{Carpinella17}

% To this end, 
% Our dependent variables are the perceived response time, presence, the impression of the agent, and user preferences. We furthermore evaluate the participants' gaze. 

% More details: 
% \begin{enumerate}
%     \item Perceived response time
%     \begin{itemize}
%         \item Appropriateness
%         \item Expectation
%     \end{itemize}
%     \item Presence
%     \begin{itemize}
%         \item Social presence
%         \item Parasocial interaction
%         \item Engagement
%         \item Social realism
%     \end{itemize}
%     \item Impression of the agent
%     \begin{itemize}
%         \item Humanlikeness
%         \item Perceived intelligence
%         \item Likeability
%         \item Willingness to interact
%         \item Competence
%         \item Naturalness
%         \item Assuredness
%     \end{itemize}
%     \item User preference
%         \begin{itemize}
%             \item Hiring choice
%             \item Impact on experience
%         \end{itemize}
% \end{enumerate}

\begin{figure}[h]
	\centering % avoid the use of \begin{center}...\end{center} and use \centering instead (more compact)
	\includegraphics[width=\linewidth]{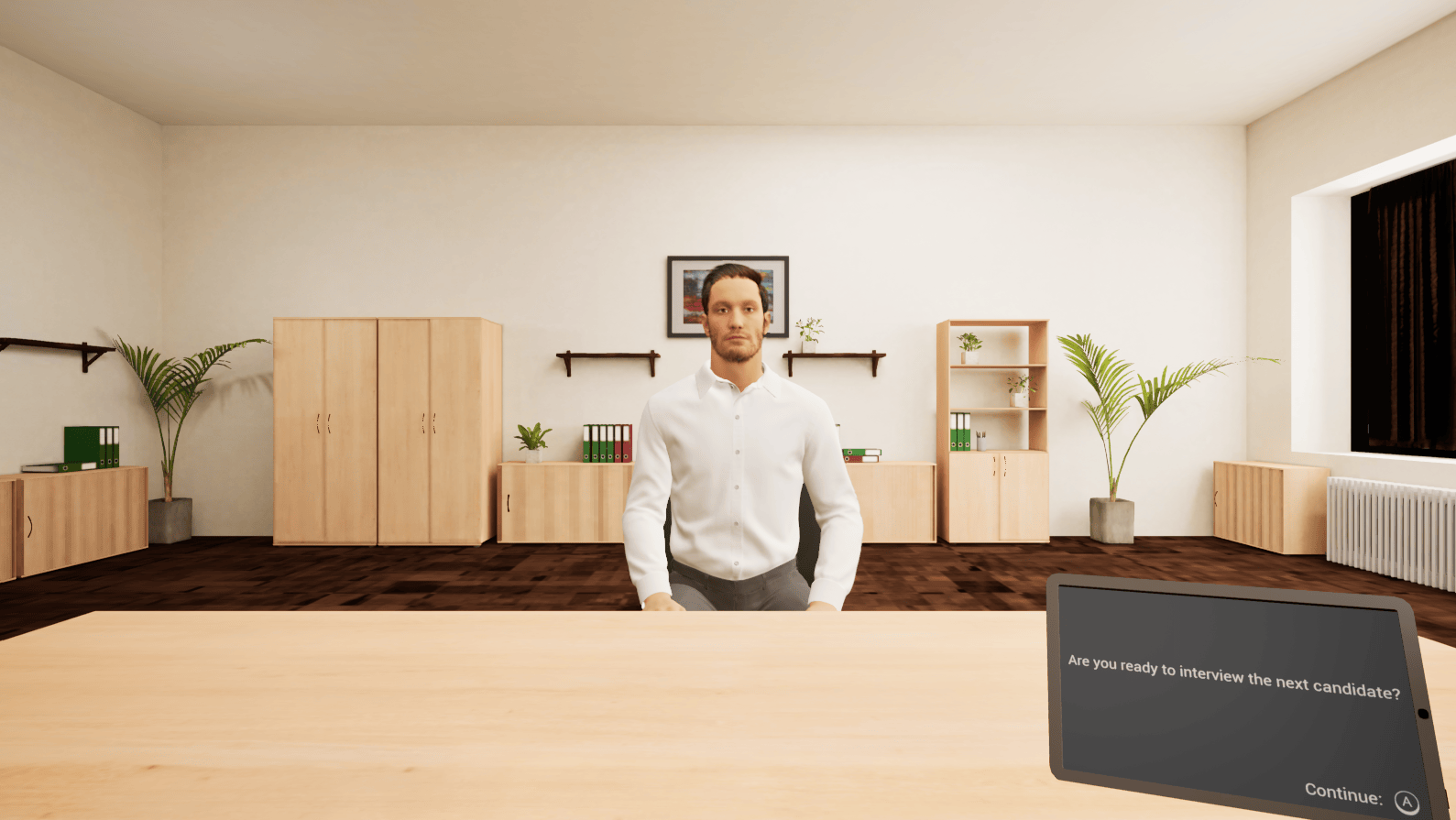}
	\caption{Setup of the virtual environment from the participant's point of view.}
	\label{fig:virtual-scene}
\end{figure}

\subsection{Procedure}
All participants were informed about the purpose and procedure of the study both verbally and in writing. They gave their informed consent by signing a consent form before participating. The study was conducted using a seated VR setup designed to enhance immersion and simulate a realistic job interview environment. The participants sat in an office chair positioned in front of a physical table that mirrored the virtual setting. During the experiment, the experimenter sat across from the participant and operated the application on a monitor facing away from the participant, allowing him to monitor the participant’s point of view throughout the session. After putting on the VR headset, participants were able to adjust the position and height of the virtual screen to match their comfort and seated posture. Eye-tracking was then calibrated using the built-in Meta Quest Pro calibration routine, followed by a brief accuracy test to verify the alignment of the participant’s gaze with the expected target points. Before starting the actual experiment, the instructor explained the interview scenario and the task to the participant.

Each session began with a short training phase conducted in the base condition, where participants asked the first two interview questions to familiarize themselves with the interaction process. After training, the scenario was restarted and the experiment began with the first interview. To control for order effects, the sequence of conditions was counterbalanced across participants. The participants then read a question aloud and pressed a button on the controller. The ECA would then enter a "thinking" phase, which varied depending on the experimental condition, before responding. This process was repeated for a total of ten questions.

Each VR interview session lasted approximately 3 minutes. After completing the second condition, eye-tracking accuracy was quickly rechecked before beginning the next scenario to ensure reliability through a separate scene (participants were asked if an object is following there gaze correctly, if not the gaze tracking was calibrated again). After each condition, participants removed the VR headset and completed a digital questionnaire on an external tablet. Filling out the questionnaire took approximately 7 minutes for each condition. After completing all four conditions, participants filled out a final questionnaire reflecting on their overall experience and their preferences across the conditions. \replaced{This was followed by a brief interview in which the experimenter asked participants to share any further comments about the time indicators and behavioral animations—or their absence—and any other feedback about their experience or the study as a whole. }{This was followed by a short interview conducted by the experimenter to collect qualitative feedback and explore participants’ impressions and reasoning behind their condition rankings}The entire study session, including setup, calibration, interviews, and questionnaires, lasted between 60 and 90 minutes. Participants were compensated with a €20 gift card for their time. The study was approved by our institutions' ethics board. 
%The questionnaire included measures for perceived response time, presence, and the impression of the virtual agent. 

\subsection{Apparatus}
The prototype was implemented using Unreal Engine 5.5.3. For the VR setup, we used the Meta Quest Pro headset connected via Meta Quest Link to a PC, enabling a PCVR configuration. The Meta Quest Pro was chosen specifically for its integrated eye-tracking functionality, which was essential for our objective measurements during the study. Participants interacted with the environment using the Meta Quest Pro controllers, where the A button served as the primary input for continuing the conversation.
The prototype application ran on a PC equipped with an NVIDIA GeForce RTX 4090 GPU and a 12th Gen Intel(R) Core(TM) i7-12700 2.10 GHz processor.
%\todo[inline]{denmar, please provide the exact processor name}

ECAs were created using Unreal Engine MetaHumans and dressed in formal business attire appropriate for a job interview scenario. We created two male and two female agents to provide diversity in the virtual interviewees while maintaining a professional and consistent appearance across all interactions. %Due to an error in randomization, each agent was shown with a consistent filler type. 
%\sj{Feel free to reformulate previous sentence.}

The virtual environment was designed to resemble a realistic office setting. It featured detailed assets such as a wooden desk, an office chair, wall-mounted shelves, filing cabinets filled with books and folders, and several potted plants to enhance immersion. The ECA was positioned in the center of the room, seated on the office chair and facing the participant. A desk separated the agent from the participant.

A virtual tablet was placed on the desk to present the interview questions. This tablet was positioned slightly to the right and angled toward the participant, allowing easy readability of the questions while avoiding obstruction of the participant’s view of the agent. The virtual setup can be seen in Fig.~\ref{fig:virtual-scene}

For the different visualization strategies, in the \textsc{embedded} condition, the time indicator was displayed as a visitor badge attached to the agent’s chest, just above the left side. In the \textsc{external} condition, the time indicator took the form of a thinking bubble that appeared next to the agent’s head on the right-hand side (from the participant’s perspective), visually resembling a thought emerging from the agent.

\subsection{Participants}
In total, 24 volunteers (17 male, 7 female, mean age: 30.83 sd: 10.79) participated in the study. Ages ranged from early 20s to late 50s. Occupations included university students (13 participants), engineers (2), research associates (2), and other roles such as programmer, software engineer, computer scientist, \replaced{homemaker}{housewife}, farmer, entrepreneur, and supplier management. All volunteers were at least 18 years old and were recruited using convenience sampling. 

Additionally, participants rated their prior experience on a 5-point Likert scale (1 = none, 5 = very experienced) across four domains: virtual reality, virtual agents, \replaced{large language models}{large-language-models}, and job interviews. The distribution of responses across these domains is summarized in Table~\ref{tab:experience_distribution}. While experience with job interviews tended to be higher, participants reported more varied levels of familiarity with VR and LLMs. In contrast, the majority indicated little to no prior experience with virtual agents. 

\begin{table}[ht]
    \centering
    \caption{Number of participants per self-rated experience level (1 = none, 5 = very experienced)}
    \setlength{\tabcolsep}{6pt}
    \begin{tabular}{|l|c|c|c|c|c|}
        \hline
        \textbf{Domain} & \textbf{1} & \textbf{2} & \textbf{3} & \textbf{4} & \textbf{5} \\
        \hline
        Virtual Reality (VR)       & 5  & 5  & 4  & 6  & 4  \\
        Virtual Agents             & 8  & 6  & 5  & 4  & 1  \\
        Large Language Models (LLMs)& 6  & 3  & 5  & 4  & 6  \\
        Job Interviews             & 2  & 4  & 9  & 5  & 4  \\
        \hline
    \end{tabular}
    \label{tab:experience_distribution}
\end{table}

\subsection{Hypotheses}
For this experiment we hypothesized that: \\
${H_1}$: Behavioral fillers will result in a more positive and natural impression of the agent and an increased sense of presence compared to symbolic fillers or the absence of fillers. \\
${H_2}$: Symbolic progress indicators will lead to an improved perception of the agent's response time compared to behavioral fillers or the absence of fillers. \\
${H_3}$: Embedded symbolic indicators will lead to a more positive and natural impression of the agent compared to an external symbolic representation. 

% SJ: H1 is based on previous work, H2 and H3 come from us and we don't have related work on symbolic progress indicators to back them up yet 

%These hypotheses are grounded in previous work. 
\added{Hypothesis ${H_1}$ is based on the results of previous studies. Kum and Lee \cite{kum2022journal} find several positive effects of gestural fillers as opposed to no fillers on the impression of the agent in a screen-based interaction. Elfleet and Chollet \cite{elfleet2024investigating} investigate multimodal fillers that include gestural behavior. Their results show, among other results, an increase in presence, anthropomorphism, and willingness to interact with the agent again.
${H_2}$ stems from work showing that people like the feedback of progress indicators when waiting for processes to finish \cite{myers1985importance}. While previous results are based on typical screen-based applications, we hypothesize that such preferences could hold true for interactions with virtual agents. 
The reasoning behind ${H_3}$ is that an external symbolic indicator is more obvious and even further from reality than an embedded indicator.}

% Elfleet and Chollet [9] investigated the effectiveness of multimodal feedback strategies (filled pauses, nonverbal turntaking behaviors, and visual feedback) in mitigating perceived latency. In a user study (n=18), they compared a baseline condition (no feedback) with a feedback-enhanced condition. Their results indicated that the multimodal feedback condition significantly improved the sense of presence and immersion. At the same time, it is important that both verbal and non-verbal feedback match [6

%${H_1}$: behavioral fillers will result in a more natural perception of the user interaction with an agent than symbolic fillers. \\
%${H_2}$: Symbolic indicators will lead to an improved perception of waiting time over behavioral fillers. \\
%${H_3}$: Embedded symbolic indicators will be seen as more natural compared to an external symbolic representation. 

%\sj{There is room for improvement here. I already changed the formulation a bit, but don't want to make big changes without telling you. We don't mention the base condition at all. Does the more natural perception of the user interaction in H1 refer to increased presence or realism, or the perception that the agent is natural, or all of them? Similar question for H3, which measures are we looking at here? We should motivate the hypotheses, for example, based on previous work.}

%\todo[inline]{JG: Yes. We should mention base - I was basically to uninspired and lack the grounding in RW. If you have the capacity to ground it a bit better, that would be much appreciated.}

%% file: sections/04_results.tex
\section{Results}

\begin{table}[t]
    \centering
    \caption{Friedman test results for the influence of \textsc{filler type} on subjective measures. d$f_1$ = d$f_{effect}$ and d$f_2$ = d$f_{error}$. Significant findings are marked in gray.}
    \setlength{\tabcolsep}{4pt}

    \begin{tabular}{|c||c|c|c|c|}
        \hline
        \multicolumn{5}{|c|}{\textbf{Perceived Response Time}} \\
        \hline
        & $\chi^2$ & d$f$ & p &  Kendall's $W$ \\
        \hline
        Appropriateness & $15.9$ & $3$ & \cellcolor{lightgray}$0.001$ & 0.221 \\
        \hline
        Expectedness & $7.18$ & $3$ & $0.066$ & 0.099 \\
        \hline                    
    \end{tabular}

    \vspace{1em}
    \begin{tabular}{|c||c|c|c|c|}
        \hline
        \multicolumn{5}{|c|}{\textbf{Presence}} \\
        \hline
        & $\chi^2$ & d$f$ & p &  Kendall's $W$ \\
        \hline
        Social Presence & $3.11$ & $3$ & $0.376$ & 0.043  \\
        \hline
        Parasocial Interaction & $10.7$ & $3$ & \cellcolor{lightgray}$0.014$ & 0.149 \\
        \hline
        Engagement & $10.3$ & $3$ & \cellcolor{lightgray}$0.016$ & 0.143 \\
        \hline
        Social Realism & $12.$ & $3$ & \cellcolor{lightgray}$0.005$ & 0.179 \\
        \hline                    
    \end{tabular}

    \vspace{1em}
    \begin{tabular}{|c||c|c|c|c|}
        \hline
        \multicolumn{5}{|c|}{\textbf{Impression of the Agent}} \\
        \hline
        & $\chi^2$ & d$f$ & p &  Kendall's $W$ \\
        \hline
        Humanlikeness & $14.6$ & $3$ & \cellcolor{lightgray}$0.002$ & 0.202 \\
        \hline
        Perceived Intelligence & $6.31$ & $3$ & $0.097$ & 0.227 \\
        \hline
        Likeability & $5.14$ & $3$ & $0.162$ & 0.071 \\
        \hline
        Willingness to Interact & $7.44$ & $3$ & $0.059$ & 0.103 \\
        \hline
        Competence & $5.48$ & $3$ & $0.14$ & 0.076 \\
        \hline
        Naturalness & $10.7$ & $3$ & \cellcolor{lightgray}$0.013$ & 0.149 \\
        \hline
        Assuredness & $5.48$ & $3$ & $0.14$ & 0.076 \\
        \hline                    
    \end{tabular}

    \label{tab:subjectiveResultsTable}
\end{table}

\begin{table}[t!]
    \centering
    %\small
    \caption{Descriptive statistics of the dependent variables, including mean and standard deviation (in parentheses). 
    RGTF: Relative Gaze Time on Face. 
    %AGDF: Average Gaze Deviation from Face. 
    Ba=\textsc{base}, Be=\textsc{behavioral}, Em=\textsc{embedded}, Ex=\textsc{external}. Measures with significant findings are marked in gray.}

    \resizebox{\columnwidth}{!}{%
    \begin{tabular}{|l|c|c|c|c|}
        \hline
        \textbf{Dependent Variable} & \textbf{Ba} & \textbf{Be} & \textbf{Em} & \textbf{Ex} \\
        \hline
    
        %AGDF & \makecell{12.33\\(14.21)} & \makecell{5.79\\(4.32)} & \makecell{9.88\\(6.98)} & \makecell{8.38\\(3.85)} \\
        %\hline
        \multicolumn{5}{|c|}{\textbf{\rule{0pt}{2.5ex}Perceived Response Time}} \\
        \hline
        
        \rowcolor{lightgray}
        Appropriateness & \makecell{4.04\\(1.94)} & \makecell{5.83\\(1.09)} & \makecell{4.46\\(1.59)} & \makecell{4.33\\(1.71)} \\
        \hline
        Expectedness & \makecell{4.33\\(1.95)} & \makecell{5.79\\(1.18)} & \makecell{4.96\\(1.65)} & \makecell{4.96\\(1.76)} \\
        \hline

        \multicolumn{5}{|c|}{\textbf{\rule{0pt}{2.5ex}Presence}} \\
        \hline
        Social Presence & \makecell{4.46\\(1.12)} & \makecell{4.80\\(1.19)} & \makecell{4.16\\(1.20)} & \makecell{4.42\\(1.12)} \\
        \hline
        \rowcolor{lightgray}
        Parasocial Interaction & \makecell{4.47\\(1.21)} & \makecell{5.05\\(1.03)} & \makecell{4.50\\(1.01)} & \makecell{4.53\\(1.13)} \\
        \hline
        \rowcolor{lightgray}
        Engagement & \makecell{4.31\\(1.23)} & \makecell{4.94\\(0.94)} & \makecell{4.24\\(1.07)} & \makecell{4.33\\(1.25)} \\
        \hline
        \rowcolor{lightgray}
        Social Realism & \makecell{4.78\\(1.32)} & \makecell{5.22\\(1.34)} & \makecell{4.47\\(1.70)} & \makecell{3.89\\(1.98)} \\
        \hline

        \multicolumn{5}{|c|}{\textbf{\rule{0pt}{2.5ex}Impression of the Agent}} \\
        \hline
        \rowcolor{lightgray}
        Humanlikeness & \makecell{3.99\\(1.27)} & \makecell{4.61\\(1.33)} & \makecell{3.53\\(1.29)} & \makecell{3.18\\(1.42)} \\
        \hline
        Perceived Intelligence & \makecell{5.19\\(0.97)} & \makecell{5.67\\(0.77)} & \makecell{5.24\\(0.97)} & \makecell{5.34\\(0.99)} \\
        \hline
        Likeability & \makecell{4.73\\(1.17)} & \makecell{5.43\\(1.01)} & \makecell{4.94\\(1.09)} & \makecell{4.90\\(1.33)} \\
        \hline
        Willingness to Interact & \makecell{4.33\\(1.86)} & \makecell{5.17\\(1.46)} & \makecell{4.54\\(1.74)} & \makecell{4.33\\(1.61)} \\
        \hline
        Competence & \makecell{4.33\\(1.79)} & \makecell{5.38\\(1.56)} & \makecell{4.54\\(1.59)} & \makecell{4.38\\(1.84)} \\
        \hline
        \rowcolor{lightgray}
        Naturalness & \makecell{3.75\\(1.33)} & \makecell{4.71\\(1.55)} & \makecell{3.67\\(1.34)} & \makecell{3.33\\(1.52)} \\
        \hline
        Assuredness & \makecell{4.33\\(1.37)} & \makecell{5.13\\(1.23)} & \makecell{4.33\\(1.37)} & \makecell{4.42\\(1.50)} \\
        \hline
        
        \multicolumn{5}{|c|}{\textbf{\rule{0pt}{2.5ex}Relative Gaze Time}} \\
        \hline
        \rowcolor{lightgray}
        RGTF & \makecell{0.66\\(0.29)} & \makecell{0.64\\(0.20)} & \makecell{0.48\\(0.25)} & \makecell{0.51\\(0.23)} \\
        \hline

        \multicolumn{5}{|c|}{\textbf{\rule{0pt}{2.5ex} Gaze on Face During ECA's Response}} \\
        \hline
        \rowcolor{lightgray}
        Before Response & \makecell{6.83\\(3.16)} & \makecell{6.67\\(2.37)} & \makecell{4.00\\(2.84)} & \makecell{4.88\\(2.86)} \\
        \hline
        \rowcolor{lightgray}
        Response Start & \makecell{6.33\\(3.25} & \makecell{8.33\\(2.51)} & \makecell{4.88\\(2.95)} & \makecell{5.67\\(2.44)} \\
        \hline
        After Response & \makecell{7.17\\(2.94)} & \makecell{8.21\\(2.52)} & \makecell{7.42\\(2.28)} & \makecell{8.21\\(2.7)} \\
        \hline
    \end{tabular}
    }

    \label{tab:descriptiveStatistics}
\end{table}

\begin{table}[!b]
    \centering
    \caption{RM-ANOVA results for the influence of \textsc{filler type} on Gaze on Face during ECA's response.  d$f_1$ = d$f_{effect}$ and d$f_2$ = d$f_{error}$. Significant findings are marked in gray.}
    \setlength{\tabcolsep}{6pt}
    
    \begin{tabular}{|l|c|c|c|c|c|}
        \hline
        \textbf{Measure} & d$f_1$ & d$f_2$ & F & p & $\eta^2_p$ \\
        \hline
        Before Response & 3 & 69 & 9.61 & \cellcolor{lightgray}$<0.001$ & 0.295 \\
        \hline
        Response Start & 3 & 69 & 11.1 & \cellcolor{lightgray}$<0.001$ & 0.325 \\
        \hline
        After Response & 2.22 & 51 & 1.73 & 0.184 & 0.070 \\
        \hline
    \end{tabular}

    \label{tab:EyeGazeOneSecond}
\end{table}

\begin{table}[!b]
    \centering
    \caption{RM-ANOVA results for the influence of \textsc{filler type} on RGTF: Relative Gaze Time on Face.  d$f_1$ = d$f_{effect}$ and d$f_2$ = d$f_{error}$. Significant findings are marked in gray.}
    \setlength{\tabcolsep}{6pt}
%AGDF: Average Gaze Deviation from Face.
    \begin{tabular}{|l|c|c|c|c|c|}
        \hline
        \textbf{Measure} & d$f_1$ & d$f_2$ & F & p & $\eta^2_p$ \\
        \hline
        RGTF & 3 & 69 & 6.42 & \cellcolor{lightgray}$<0.001$ & 0.218 \\
       % AGDF & 1.54 & 35.51 & 3.47 & 0.053 & 0.131 \\
        \hline
    \end{tabular}

    \label{tab:RGTFResultsTable}
\end{table}

We collected the relative gaze time on the face, i.e., the time the participants looked at the face in relation to the overall time, during the ECA's thinking phase (the time between the end of the participant asking a question and the agent's response).
We analyzed the influence of \textsc{filler type} on this measure using one-way repeated measures analysis of variance (RM-ANOVA). 
%, with Greenhouse-Geiser correction if the sphericity assumption was violated.
Post-hoc comparisons were conducted using pairwise t-tests with Tukey correction for multiple comparisons, controlling the error rate at an initial significance level of $\alpha = 0.05$. Effect sizes for the statistical tests were calculated using Partial Eta Squared $\eta^2_p$ and for the post hoc comparisons were computed using Cohen's $d_z$.

We collected the subjective measures for perceived response time, presence, and the impression of the agent using questionnaires.
For these, we used Friedman tests for one-way repeated measures analysis of variance by ranks, as the data were collected using a 7-point Likert scale. The results of the statistical tests for the subjective measures are presented in Table \ref{tab:subjectiveResultsTable}.
Post-hoc pairwise comparisons were performed using the Durbin-Conover test with Holm correction. Agreement scores for the Friedman test were computed using Kendall's $W$, and effect sizes for the post hoc comparisons were computed using Pearson's $r$.
The descriptive statistics for all the dependent variables are presented in Table \ref{tab:descriptiveStatistics}. 
To aid replication, the data underlying the statistical analysis available on \href{https://osf.io/2gej9/?view_only=3600ee12cfa64fb9b8527bb86d40a4f4}{OSF} and the code base for the experiment is available on  \href{https://gitlab.com/mixedrealitylab/eca-delay-mitigation-fillers-vr#}{GitLab}.

\begin{figure*}[t]
	\centering % avoid the use of \begin{center}...\end{center} and use \centering instead (more compact)
	\includegraphics[width=\linewidth]{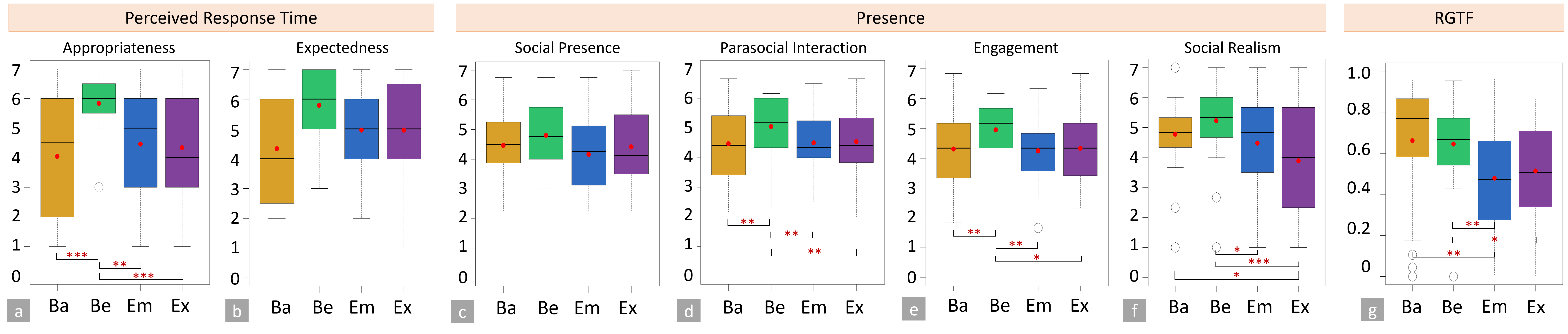}
	\caption{
        Box plots for the measures perceived response time, presence, and RGTF: a) Appropriateness. b) Expectedness. c) Social Presence. d) Parasocial Interaction. e) Engagement. f) Social Realism. g) RGTF (measured in percentage). Ba=\textsc{base}, Be=\textsc{behavioral}, Em=\textsc{embedded}, Ex=\textsc{external}. 
         The number of stars indicates the significance levels between the techniques: *** \textless 0.001, **  \textless 0.01, *  \textless 0.05. %Equivalence between conditions is indicated by $\equiv$.
        }
	\label{fig:perceivedRT-presence-plots}
\end{figure*}

\subsection{Perceived Response Time} 

Perceived response time was collected across two dimensions, appropriateness and expectedness.
%\paragraph{Appropriateness.}
We found a significant effect of \textsc{filler type} on appropriateness. Post hoc tests revealed a significantly higher appropriateness score in \textsc{behavioral} compared to all other other conditions \textsc{base} ($ p < 0.001 $, $r = 0.81 $), \textsc{embedded} ($ p = 0.002 $, $r = 0.67 $), and \textsc{external} ($ p < 0.001 $, $r = 0.71 $), see Fig. \ref{fig:perceivedRT-presence-plots}(a).

%\paragraph{Expectedness.}
However, we did not find a significant effect of \textsc{filler type} on expectedness in the perceived response time. 
\added{To further analyze expectedness, we investigated the participants' eye gaze behavior, specifically, where exactly the participants looked one second before the response started, at the start of the response, and one second after. We used a threshold of one second, as the shortest possible delay in our use case was two seconds. We summed up the number of questions out of all ten, in each condition, where the participants looked at the agent’s face in each condition. %We conducted a one-way RM-ANOVA for these analyses, as the dependent variable in this case, precisely, the number of questions in which participants looked at the agent’s face, was continuous. We used Greenhouse-Geisser correction before RM-ANOVA when sphericity was violated (c.f. \cite{blanca2023non}). Post-hoc comparisons were conducted using pairwise t-tests with Tukey correction for multiple comparisons, controlling the error rate at an initial significance level of $\alpha = 0.05$. 
}

\added{An RM-ANOVA indicated a significant effect of \textsc{filler type} on gaze on face one second before the response started, with post-hoc tests indicating significant differences between \textsc{base} and \textsc{embedded} ($ p=0.001$, Cohen's $d_z=0.89$), \textsc{behavioral} and \textsc{embedded} ($p<0.001$, Cohen's $d_z=0.98$), and \textsc{behavioral} and \textsc{external} ($p=0.004$, Cohen's $d_z=0.79$).
We also found a significant effect of \textsc{filler type} on gaze on face right at the response start, with post-hoc tests indicating that the gaze on face was significantly higher in \textsc{behavioral} than \textsc{base} ($p=0.014$, Cohen's $d_z=0.69$), \textsc{embedded} ($p<0.001$, Cohen's $d_z=1.37$), and \textsc{external} ($p<0.001$, Cohen's $d_z=1.02$). 
The \textsc{filler type} did not have a significant effect on the users' gaze on face one second after the response started. This is supported by the means showing that for most questions, the gaze was indeed on the face after the response started, in all conditions, \textsc{base} ($m=7.17$, $sd=2.94$), \textsc{behavioral} ($m=8.21$, $sd=2.52$), \textsc{embedded} ($m=7.42$, $sd=2.28$), and \textsc{external} ($m=8.21$, $sd=2.7$). The results of the statistical tests are also shown in Table \ref{tab:EyeGazeOneSecond} and the descriptive statistics are shown in Table \ref{tab:descriptiveStatistics}.}

\added{This eye-gaze behavior suggests that participants expected the response more naturally with the \textsc{behavioral} fillers. This is shown by their increased visual attention to the agent’s face just before and at the start of the response. This aligns with the idea of expectedness that when users have a feeling that a response is about to happen, they naturally focus their visual attention on the speaker's face. So, this eye-gaze data provides an indirect way of understanding expectedness, despite participants’ self-reported ratings not showing a clear difference. We discuss this interpretation further in Section \ref{user-feedback-and-discussion}.}

\subsection{Presence} 

For the presence measures, we collected data for social presence, parasocial interaction, engagement, and social realism.
%\paragraph{Social Presence.} 
We did not find a significant effect of \textsc{filler type} on social presence scores.

\paragraph{Parasocial Interaction.} We found a significant effect of \textsc{filler type} on parasocial interaction scores. Post hoc tests revealed a significantly higher score in \textsc{behavioral} than all other conditions, \textsc{base} ($ p = 0.006 $, $r = 0.58 $), \textsc{embedded} ($ p = 0.007 $, $r = 0.57 $) and \textsc{external} ($ p = 0.006 $, $r = 0.58 $), see Fig. \ref{fig:perceivedRT-presence-plots}(d).

\paragraph{Engagement.} We found a significant effect of \textsc{filler type} on engagement scores. Post hoc tests revealed a significantly higher score in \textsc{behavioral} than in all other conditions, \textsc{base} ($ p = 0.004 $, $r = 0.62 $), \textsc{embedded} ($ p = 0.01 $, $r = 0.54 $), and \textsc{external} ($ p = 0.013 $, $r = 0.52 $), see Fig. \ref{fig:perceivedRT-presence-plots}(e).

\paragraph{Social Realism.} We also found a significant effect of \textsc{filler type} on social realism. Post hoc tests revealed a significantly higher score in \textsc{behavioral} than in \textsc{embedded} ($ p = 0.036 $, $r = 0.44 $) as well as \textsc{external} ($ p < 0.001 $, $r = 0.78 $). Additionally, \textsc{base} resulted in significantly higher score than \textsc{external} ($ p = 0.026 $, $r = 0.46 $), see Fig. \ref{fig:perceivedRT-presence-plots}(f).

\subsection{Impression of the Agent} 
\begin{figure*}[ht!]
	\centering % avoid the use of \begin{center}...\end{center} and use \centering instead (more compact)
	\includegraphics[width=\linewidth]{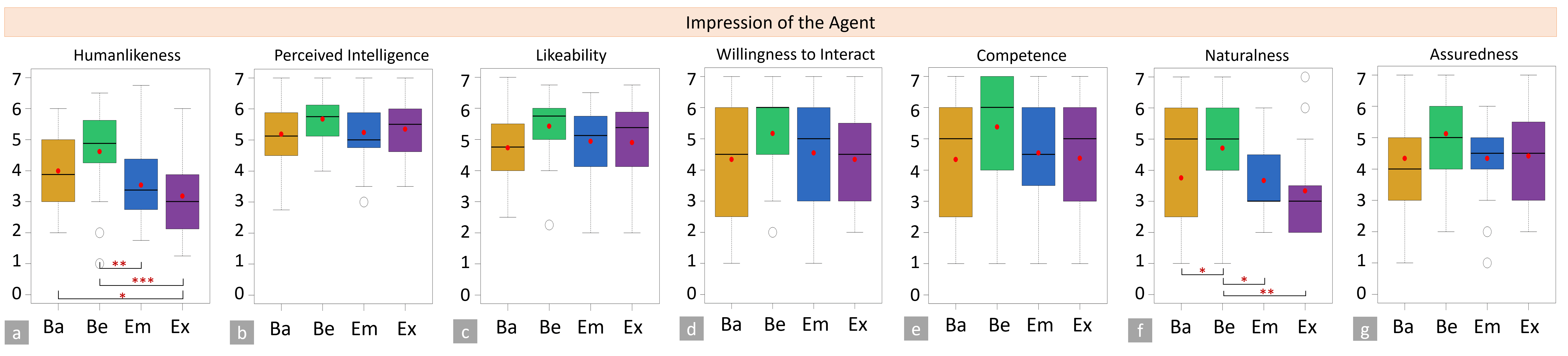}
	\caption{
        Box plots for impression of the agent: a) Humalikeness. b) Perceived Intelligence. c) Likeability. d) Willingness to Interact. e) Competence. f) Naturalness. g) Assuredness. Ba=\textsc{base}, Be=\textsc{behavioral}, Em=\textsc{embedded}, Ex=\textsc{external}. 
         The number of stars indicates the significance levels between the techniques: *** \textless 0.001, **  \textless 0.01, *  \textless 0.05. %Equivalence between conditions is indicated by $\equiv$.
        }
	\label{fig:impression-plots}
\end{figure*}

We found significant effects of \textsc{filler type} on humanlikeness and naturalness. Post hoc tests revealed a significantly higher humanlikeness score in \textsc{behavioral} than in \textsc{embedded} ($ p = 0.007 $, $r = 0.56 $) and \textsc{external} ($ p < 0.001 $, $r = 0.83 $). 
Furthermore, the humanlikeness score in \textsc{base} was significantly higher than \textsc{external} ($ p = 0.036 $, $r = 0.44 $), see Fig. \ref{fig:impression-plots}, (a).
Post hoc tests further revealed a significantly higher naturalness score in \textsc{behavioral} than in all other conditions,  \textsc{base} ($ p = 0.045 $, $r = 0.42 $), \textsc{embedded} ($ p = 0.033 $, $r = 0.44 $), and \textsc{external} ($ p = 0.001 $, $r = 0.7 $), see Fig. \ref{fig:impression-plots}, (f).

No significant differences of \textsc{filler type} were found on perceived intelligence, likeability, willingness to interact, competence, or assuredness.

\subsection{Relative Gaze Time on Face} 
Relative gaze time on the face was not normally distributed, but the sphericity assumption was met. Hence, we used the parametric, one-way RM-ANOVA, as normality violations do not tend to have a major impact on the robustness of the analysis, when sphericity is met \cite{blanca2023non}. The result is presented in Table \ref{tab:RGTFResultsTable}.

We found a significant effect of \textsc{filler type} on the relative eye-gaze time spent looking at the virtual agent's face, see \cref{fig:perceivedRT-presence-plots}, (g). Post hoc tests revealed a significantly longer relative gaze time in \textsc{behavioral} compared to \textsc{embedded} ($ p = 0.008 $, Cohen's $d_z = 0.73 $), as well as \textsc{external} ($ p = 0.024 $, Cohen's $d_z = 0.63 $). Additionally, \textsc{base} resulted in a significantly longer gaze time than \textsc{embedded} ($ p = 0.007 $, Cohen's $d_z = 0.75$).

Further, in the \textsc{embedded} condition, the participants looked at the badge on average 20.7\% of the time (sd = 18) and on the face 48\% of the time (sd = 25). In the \textsc{external} condition, the participants looked at the thinking bubble on average 32.2\% of the time (sd = 17.1) and on the face 51\% of the time (sd = 23).

\subsection{Preferences}
Participants were asked about their preferences after experiencing each condition. 66.7\% of participants (16 out of 24) preferred the \textsc{behavioral} condition, 12.5\% preferred the \textsc{embedded} (Badge) condition, and 4.2\% preferred the \textsc{external} (Bubble) condition. Only 8.3\% of participants preferred the \textsc{base} condition without any delay indicators or animations.

%% file: sections/05_discussion.tex
\section{User Feedback and Discussion}
\label{user-feedback-and-discussion}
When examining the measures for presence, we found that the \textsc{behavioral} condition resulted in significantly higher parasocial interaction, engagement, and social realism compared to all other conditions. \added{There was no significant effect for social presence. A closer look at the individual questions of this measure indicated that two of the four questions hold a similar pattern -- \textsc{behavioral} was rated significantly higher than some of the other conditions -- whereas there were no significant effects for the other two questions (about similarity to the face-to-face meeting and being in the same room). It appears that all conditions were rated similarly for these two questions. Still, with four of the five measures on presence being in line and the fifth not showing any contradicting trends, these results clearly support the first part of $H_1$ that behavioral fillers result in an increased sense of presence.}

In addition, \textsc{behavioral} also led to significantly higher humanlikeness, and naturalness ratings compared to \textsc{embedded} and \textsc{external}. %\added{There was no significant effect for social presence.}
\added{There were no significant differences for the other five measures examining the impression of the agent (perceived intelligence, likeability, willingness to interact, competence and assuredness). We noticed that the \textsc{behavioral} condition had the highest mean. %and that the p-values were low (between 0.059 and 0.17) for each measure (Figure \ref{fig:impression-plots} and Table \ref{tab:subjectiveResultsTable}).
One possible interpretation of the results would be that differences might have lower effect sizes than anticipated and therefore did not lead to significant results in our study. While our results do not allow any clear conclusions in that regard, there are no indications in any measures that could indicate a negative impression of \textsc{behavioral} compared to the three other conditions. Taken together, we conclude that $H_1$ can be supported, meaning that behavioral fillers can result in a more positive and natural impression of the agent compared to symbolic fillers or the absence of fillers.}

\added{Our results on the effect of the behavioral fillers on presence and the impression of the agent are mostly in line with Kum and Lee's and Elfleet and Chollet's results \cite{elfleet2024investigating, kum2022journal}. There are differences, however, for example, }we did not find a significant increase in the willingness to interact with the agent in the future. It is possible that visual indicators have beneficial effects when added on top of behavioral fillers. \added{Further differences in the experiment design, such as the agent being visible on a screen in Kum and Lee's study, might lead to the perceived differences. Finally, another reason could be that the tasks were not the same.}

\deleted{Our findings show that the \textsc{behavioral} condition leads to a higher perceived appropriateness of the response time, higher parasocial interaction, engagement, social realism, humanlikeness and naturalness than the \textsc{base} condition is largely in line with the results from Elfleet and Chollet \cite{elfleet2024investigating}. }
\deleted{We can conclude that the behavioral animations that we used without a visual feedback indicator are sufficient to improve the users’ experience.}

\added{Our results furthermore }indicated that the response time was perceived as significantly more appropriate in \textsc{behavioral} compared to all other conditions, which is in line with Kum and Lee's and Elfleet and Chollet's results \cite{elfleet2024investigating,kum2022journal}. Contrary to our expectations, we could not find a difference with regard to expectedness, not confirming our hypothesis $H_2$. It seems \deleted{either} that the symbolic time indicators did not significantly help participants in assessing when to expect an answer from the virtual agent for the provided range of delays (2-8 seconds). 
\added{Additionally, our findings from eye-tracking analysis imply that, with the behavioral filler, users focus their visual attention on the agent’s face and perceive the response as more natural, as it follows directly from the ending of the agent’s thinking animation and is supported by user feedback, for example, as noted by P6 regarding the behavioral filler: "It felt as if there was no delay at all."
%, as supported by the user feedback like "felt as if there was no delay at all” (P6). 
In contrast, symbolic progress indicators may divide users’ attention between the agent’s face and the visual progress indicators. If users are not focused on the agent at the moment the response begins, it may feel abrupt or “unexpected”, reducing the perception that the response arrived at the right time.}
\added{Another reason for not finding higher expectedness in the conditions with the progress indicators might be the wording of the “expectedness” question: “The virtual human you just talked to answered when expected.” We designed this statement to assess whether participants could predict the agent’s response time based on the fillers shown. However, its interpretation was left to the participants, which may have led to varying misinterpretations, e.g., interpreting it as the delay for answering that specific question in a real-world context. This ambiguity could be one of the reasons why the hypothesized differences in $H_2$ were not observed. Future work can refine and validate this measure to capture expectedness in ECA response time more precisely. }

\added{Finally, we }did not find significant differences between \textsc{embedded} and \textsc{external} for any of the measures, not confirming our hypothesis $H_3$ \added{that embedded symbolic indicators will lead to a more positive and natural impression of the agent compared to external symbolic indicators. It is possible that the indicators lead to enough awareness on the agent being an agent that the embedding did not have any further effects.}

The participants' preferences and feedback also correlate with the collected subjective metrics. 
\added{We analyzed the interview results using thematic analysis. We organized the feedback corresponding to the four experimental conditions and further classified the responses within each category as either positive or negative points of the condition. }

Participants who preferred the \textsc{behavioral} condition mentioned that "it made the agent seem more realistic" (P1, P3, P7), involved natural gestures and facial expressions (P21), and felt as "if there was no delay at all" (P6). P12 noted that "time went by faster", and P24 said they could "follow the agent’s thinking process". 
\deleted{Several participants stated that the symbolic indicators felt unnatural or unrealistic (P8, P10), and P22 mentioned they reminded them that "they were in a simulation".}
\added{Only one participant (P4) reported not feeling immersed in the \textsc{behavioral} condition, attributing this to their awareness of the simulated nature of the interaction. This was reflected in their near-zero RGTF score (see Fig. \ref{fig:perceivedRT-presence-plots}), marking them as an outlier. However, such behavior was not observed in other participants.}

Participants who preferred the \textsc{embedded} condition appreciated that it let them "know when the agent would answer" (P11, P16) and found the badge minimalistic and preferred over the bubble (P5). One participant said the badge helped indicate "whether the agent had understood the question" (P16).

Only one participant (P13) clearly preferred the \textsc{external} condition. They mentioned that it showed "the program wasn’t lagging" and helped indicate how long the agent needed to respond. Another participant (P4), although not preferring \textsc{external}, appreciated that it confirmed the system was working, but others found it "cartoonish" or "distracting" (e.g., P3, P7, P14, P18). \added{Several participants stated that the symbolic indicators felt unnatural or unrealistic (P8, P10), and P22 mentioned they reminded them that "they were in a simulation".}

Two participants (P17, P19) preferred the \textsc{base} condition, stating that the absence of indicators felt "more natural" and that the "bubble destroyed immersion". However, several others who did not prefer \textsc{base} described it as confusing, as they were unsure whether the system had registered their input (P6, P8, P12).
\added{Furthermore, three participants (P4, P13, P17) showed near-zero RGTF in the \textsc{base} condition and were outliers for this measure, frequently looking around the environment, likely due to the lack of feedback. These observations suggest that the \textsc{base} condition, which relies solely on idle animations, may confuse the user and cause potential annoyance in the absence of responsive system feedback.}

Also, in \textsc{external} and \textsc{embedded}\added{,} participants looked at the agent's face for significantly less time than in \textsc{behavioral}, and for \textsc{embedded} also significantly shorter than \textsc{base}. This effect is expected and seems plausible as participants likely diverted their gaze to the progress indicators outside the face.

Interpreting the results, we see that the \textsc{behavioral} visualization was clearly preferred in the setting for its natural behavior and that participants—even though they knew that they were talking to an agent disliked the symbolic indicators for breaking immersion. Still, it was acknowledged that visual symbolic indicators have the advantage of giving the user feedback on the estimated waiting time. It also reminds the user that the conversational partner is an agent and not another human being, which might be considered as an advantage or not, depending on the application. A potential downside of this visualization is its dependency on the actual virtual agent appearance. For example, a badge might fit in a virtual office environment as used in the experiment but not in an adventure scenario in a virtual forest. In general, it should be explored further if the results of our experiment would generalize to other settings, including different scenarios, e.g., in virtual nature environments, multi-agent environments, or game-like settings. Specifically, for the \textsc{embedded} visualization, content creators could have many liberties in blending it with the agent's appearance (e.g., embedding it into the forehead or eyes of an agent with a robot-like appearance).

\added{Different filler types may be appropriate for different use cases, e.g., behavioral fillers may be more effective in immersive scenarios where embodiment and naturalness are prioritized, such as ours, while symbolic indicators might be more suitable in time-sensitive applications (e.g., receiving AI assistance in urgent tasks), where clear information about the delay duration is important. Future research can explore the differences among various filler types in relation to specific use cases.}
\added{Additionally, during substantially longer delays, behavioral fillers might become less effective over time as they do not give users any feedback on when the agent will respond. However, the symbolic fillers offer progress indication, which may help maintain user engagement. Future research can replicate our experiment with longer delays and evaluate the impacts of the various fillers.}
\added{In our user study, we did not investigate the impact of different filler types across the individual delay durations of 2, 4, and 8 seconds, as these were not analyzed separately. Future research could evaluate how filler types interact with varying delay durations, with delay as an independent variable. }

We did not add symbolic indicators on top of behavioral fillers in addition to assessing their effect individually, which could have added further insights about their effects. 
Furthermore, we did not entangle the influence of the visual animations in the \textsc{behavioral} condition from the effect of the non-lexical fillers. Specifically, it needs to be determined in future work if the results would have been comparable if the non-lexical fillers would have been omitted and only the animations had been played.

One limitation of our study is that each filler type was always displayed with the same character. While the four characters look very similar and inconspicuous and are all wearing similar clothes, we can not exclude that character appearance might have influenced our results.
\added{Additionally, an idle animation was looped in \textsc{base}, \textsc{embedded} and \textsc{external}, along with their respective visual indicators. This idle animation consisted of natural eye movement behaviors like blinking, brief gaze shifts, and minor movements to prevent the agent from appearing frozen and showing humanlikeness. This resulted in the agent occasionally diverting its gaze shortly after the participant finished asking a question, which may have conveyed unintended cues about the agent's responsiveness.}

Another limitation could be the specific use case and the setting of our experiment. The task given to the participants was to interview the virtual agent, which could influence their expectations of getting a human-like response. However, other tasks that solely focus on information acquisition from LLMs could benefit from the other filler types. 
\added{We speculate that for other ECA tasks where information retrieval is the main goal, the symbolic fillers could impact user experience more positively than behavioral fillers.} Future work can explore alternative use cases where information retrieval is the main goal and assess the impact of various filler types on user preferences.

%\added{In this work, we explored symbolic progress indicators as two of the filler types, with \textsc{embedded} positioned closer to the ECA's face. Here, we used the agent’s badge as a neutral visual element that was compatible across genders. However, this choice was very specific to the job interview scenario, which may not hold in other settings. We believe that in the cases of non-human like ECAs, the design space for embedded fillers would be much larger, and these areas could also be interesting venues for future research.} 

\added{In our user study, we used pre-defined delays in the response time that allowed a smooth transition from the behavioral filler animation sequences to the response states. However, in real-world applications, system response times are often unpredictable, and deploying the behavioral fillers effectively can be challenging, e.g., the response has already arrived, but the agent is in mid-animation. Hence, certain trade-offs must be considered to balance the response time and the perceived naturalness in the case of behavioral fillers.
Prior work from Gnewuch et al. \cite{gnewuch2018faster} demonstrated that chatbots using dynamically delayed responses were perceived as more human-like and socially present, leading to greater user satisfaction than chatbots that responded instantly. Therefore, delaying an arrived response to allow the behavior filler to complete would help in keeping the perceived naturalness intact, however, it is important to do this optimally.
Instead of fixed-length animations, an animation state machine could play small modular filler animations until the response has arrived. Once a response is ready, the system can play a short finisher animation (e.g., a nod or an "Ah!" expression) before presenting it, effectively balancing both naturalness and responsiveness.}

\added{Similarly, it is important to visualize the estimated waiting times appropriately in the progress bar, as these visualizations could be inaccurate due to variance in response time. To mitigate this, user experience designers employ strategies like non-linear animation patterns (such as fast-to-slow fills) \cite{villar2013meta} or integrate adaptive time prediction algorithms to dynamically estimate remaining work \cite{coppa2015data}. Future work can explore the aforementioned animation state machines and the strategies for progress indications for behavioral and symbolic fillers, respectively, in real-time conversational scenarios. Future work can also investigate the impact of explicitly showing estimated waiting time on the perceived response time.}

%The behavioral animation is the best condition we could create

%\subsection{Limitations}

%There are several limitations to our work.

%% file: sections/06_conclusion.tex
\section{Conclusion}

In this work, we explored behavioral and symbolic fillers as delay mitigation strategies when communicating with embodied conversational agents in virtual reality. While the multimodal behavioral fillers included thinking animations with non-lexical verbal fillers, symbolic fillers included progress indicators either integrated into a badge on the ECA or in a thinking bubble. 

Our results indicated that the multimodal behavioral filler increased the appropriateness of the perceived response time, lead to higher perceived aspects of presence (parasocial interaction, engagement, and social realism), and improved the impression participants had of the agent when it came to humanlikeness, and naturalness. 
Our results also suggest that symbolic time indicators can not easily replace behavioral delay mitigation strategies for ECAs. In our scenario, they did not seem to substantially better  help participants in judging when to expect an answer from the ECA. Even if individual participants preferred our symbolic time indicator conditions, as it confirmed that the system was working and indicated when to expect an answer, the majority of participants preferred behavioral delay mitigation strategies. 

While these results do not speak in favor of symbolic time indicators, we did not explore if adding symbolic indicators on top of behavioral fillers would be considered useful by users. Future work will need to investigate if mixing such different delay mitigation strategies might be advantageous. 

%Adding symbolic indicators on top of behavioral mitigation strategies might still have advantages and should be investigated. 

%Using important keywords of the prompt/query to show along with progress bar, to indicate as a feedback what the virtual agent is thinking/processing.